\documentclass[12pt,preprint]{aastex}
\newcommand{\beq}{\begin{equation}}
\newcommand{\eeq}{\end{equation}}
\newcommand{\ben}{\begin{eqnarray}}
\newcommand{\een}{\end{eqnarray}}
\newcommand{\bth}{\vec{\theta}}
\newcommand{\hbth}{\hat{\theta}}
\newcommand{\bu}{{\bf u}}

\begin{document}
\title{Reconstructing the Peak Distribution in the Sunyaev-Zeldovich
Effect Surveys} 
\author{\sc Jounghun Lee\altaffilmark{1} and
Naoki Yoshida\altaffilmark{2}}
\altaffiltext{1}{Department of Physics, The University of Tokyo, Tokyo 
113-0033, Japan ; lee@utap.phys.s.u-tokyo.ac.jp}
\altaffiltext{2}{National Astronomical Observatory Japan, Mitaka, 
Tokyo 181-8588, Japan ; naoki@th.nao.ac.jp}

\received{2003 Nov. 1}
\accepted{2003 ???}
\begin{abstract} 
We examine the ability of the Sunyaev-Zel'dovich effect (SZE) 
statistics proposed by Lee (2002) using numerical simulations.
The statistics describe the distribution of the peak heights
in the noise-contaminated SZE sky maps, and provide an analysis
technique to sort out the noise-contamination effect and estimate 
the number density of real clusters. 
The method is devised to be suitable for the interferometric 
SZE observations in drift-scanning mode like AMiBA experiment. 
We apply the proposed method to a set of realistic SZE sky maps 
constructed from large-scale cosmological simulations, and show that 
the method indeed allows us to estimate the number density of clusters 
efficiently. The efficiency of the method is demonstrated in two 
aspects: (i) it can reconstruct the number density of 
clusters even though the clusters contribute only a small fraction 
($\sim 10 \%$) of the total number of peaks in each SZE map; 
(ii) it can count even those clusters whose amplitudes are low enough 
to be compatible with the noise level while conventional method of 
cluster identification using the threshold cutoff cannot count them 
properly within the same observation time.  
Thus, the proposed method may be useful for the study of 
cluster abundance exploiting future interferometric SZE surveys 
such as AMiBA.
\end{abstract} 
\keywords{cosmology: theory  --- SZ effect: clusters}

\section{INTRODUCTION}

The thermal Sunyaev-Zel'dovich effect (SZE) by galaxy clusters represents 
the spectral distortion of the cosmic microwave background radiation 
(CMBR) through the inverse Compton scattering by the hot intra-cluster 
electrons \citep{sun-zel72}.  Since the SZE is independent of the redshift, 
it provides a unique tool to detect high-redshift clusters. Thus, the 
blank-field SZE survey can measure the formation epoch and the abundance
evolution of galaxy clusters, which can be used as a probe to put constraints 
on the cosmological parameters 
\citep{bar-etal96,hen97,bah-fan98,via-lid99,bor-etal99,fan-chi01,
gre-etal01,mol-etal02}. 

The SZE flux diminution is directly proportional to the cluster 
Comptonization parameter $y$, defined as 
\beq
y(\bth) = \frac{\sigma_{T}k_{B}}{m_{e}c^{2}}
\int n_{e}(l\hbth)T_{e}(l\hbth)dl, 
\label{eqn:compton} 
\eeq 
where $\bth$ represents the location vector along the line-of-sight 
($\hbth \equiv \bth/\vert \bth\vert$) and the integral is over the 
CMBR path in the line-of-sight direction. Hence the SZE survey at a fixed 
frequency basically provides a map of the $y$-parameter. The number
density of galaxy clusters, however, cannot be directly measured from
the observed SZE maps because of the presence of noise. 

Conventionally, signals from the noise-contaminated maps are 
identified as the local peaks whose flux amplitudes (for the SZE
observation, proportional to the peak height of the Comptonization
parameter, $y$) 
significantly exceed the noise level. Therefore, to detect lower-amplitude 
signals using the conventional approach, one has to decrease the noise 
level by increasing the observational time. Unfortunately, the necessary 
observational time increases inversely proportional to the squares of the 
imposed noise level. It is clearly inefficient to use the conventional 
signal-selection method for the estimate of the cluster number counts,  
because of the high costs of the interferometric observations.  

Recently \citet{lee02} proposed a statistical strategy to estimate efficiently
the number density of galaxy clusters from the noise-contaminated SZE 
maps constructed from the interferometric observation in drift-scanning
mode, with the AMiBA experiment as a target model.  
By {\it the efficient estimate},  we mean that the method is
capable of estimating the number density of clusters which contribute 
only a small fraction of the total number density of peaks in the 
SZE maps, counting properly even those clusters whose peak heights 
are low enough to be compatible with the noise level without increasing 
the observational time. 
The key feature of the method is to exploit the property that the 
noise-contamination is described by a Gaussian statistic.
Then the cluster number density can be estimated by simply deconvolving 
a one-dimensional Gaussian function and the zero-th order approximation 
to the cluster number density.  Although the method has been shown to 
render reasonable results for idealized Monte-Carlo simulations 
\citep{lee02}, it is necessary to test its ability by applying 
it to more realistic simulations before using it in real SZE surveys.  

In this {\it Letter}, we test the ability of the SZE statistical strategy 
using the outputs of a large cosmological simulation. 
In $\S 2$, we provide a concise overview of the model experiment AMiBA
and the SZE statistics. In $\S 3$, we describe the numerical simulation, 
explain how to apply the statistical strategy to the simulated SZE maps, 
and compare the reconstructed cluster number densities with the original 
ones. In $\S 4$, we summarize the results, and give concluding remarks.

\section{The SZE CLUSTER SURVEY}

\subsection{AMiBA in a Drift-Scan Mode}

For the blank-field SZE survey,  we have two types of observations: 
the single dish bolometric observations such as BOLOCAM 
\citep{gle-etal98} and the multiple dish interferometric 
observations such as our model experiment, AMiBA \citep{lo-etal00}. 
AMiBA (Array for Microwave Background Anisotropy) is an interferometer 
telescope with $19$ dishes of $1.2$ meter,  dedicated to SZE 
observation at operating frequency of $90 \pm 16$ GHz. 
It is designed to be most suitable for a deep survey of low-mass
clusters at high redshifts. For this purpose, it has relatively small 
beam of $2$ arcminute, and plans to survey $5$ square degree for 20
hours per field  on average (the total period is over $7$ months), 
expected to be able to detect $1$ cluster per every $9$ hour.  

An observational strategy of AMiBA is to use the drift-scan method,  
which will help keep the ground fringes constant with time, improve the 
Fourier-space resolution, and most importantly simplify the noise
analysis \citep{pen-etal02}. A drift-scanned sky map from AMiBA experiment 
will be a combination of a small fraction of cluster signals 
($\sim 10\%$) and a large fraction of dominant instrumental noise.  
The distribution of cluster signals is non-Gaussian, while that of the
instrumental noise is Gaussian with no spatial correlation. 
Since the statistical property of a Gaussian field is well known, 
the noise analysis for the SZE sky maps from AMiBA experiment 
in drift-scanning mode may be analytically tractable 
\citep{zha-etal02,lee02}.  

\subsection{The SZE Statistics : An Overview}
 
The peak number density of a Gaussian random field is analytically given by 
\citep{lon57,bon-efs87}: 
\beq
n_{g}(\nu) = \frac{1}{(2\pi)^{3/2}}R_{*}^{-2}e^{-\frac{\nu^2}{2}}
\int^{\infty}_{0} [x^2+e^{-x^2}-1]
\frac{\exp[-\frac{1}{2}(x-\gamma\nu)^{2}/(1-\gamma^{2})]}
{[2\pi(1-\gamma^2)]^{1/2}}dx,   
\label{eqn:ng}
\eeq 
where $\nu$ is the peak height rescaled by the noise 
dispersion, $\sigma_{0}$, 
$\sigma^{2}_{i} = \frac{1}{(2\pi)^{2}}\int \vert 
W(u)\vert^{2}u^{2i}d^{2}{\bf u}$ with a  noise filter $W(u)$,  
$R_{*} \equiv \sqrt{2}(\sigma_{1}/\sigma_{2})$, and 
$\gamma \equiv \sigma_{1}^{2}/(\sigma_{0}\sigma_{2})$. 
Equation (\ref{eqn:ng}) implies that the number density of the Gaussian
noise is a sharply decreasing function of $\nu$, which in turn suggests 
that the noise effect will be almost completely negligible for very 
high signals $\nu > 5$. 

The presence of noise contaminates the SZE sky map, making it obscure 
to count the number density of true signal peaks in many different ways: 
it changes the amplitudes and locations of the signal peaks, and also it 
creates or compromises signal peaks. To make matters worse, the 
fraction of the signal peaks is usually only a few ten percent of the noise
peaks.  \citet{lee02} suggested, however, that all the various 
noise-contamination effects on the number density of real clusters 
$n_{cl}(y)$ could be quantified by a one-dimensional Gaussian
distribution with the unit dispersion $p(x) =
(1/\sqrt{2\pi})e^{-x^2/2}$, and that $n_{cl}(y)$ be statistically
estimated by deconvolving $p(x)$ and the zeroth-order approximation to 
the cluster number density $n^{(0)}_{cl}(y) \equiv n_{sz}(y) - n_{g}(y)$ 
where $n_{sz}(y)$ is the number density of local peaks in a SZE map:  
\beq 
n_{cl}^{(0)}(y) = \frac{N_{cl}^{\rm tot}}{N_{sz}^{\rm tot}}
\int p(x)n_{cl}(y-x)dx,
\label{eqn:dec}
\eeq  
where $N_{sz}^{\rm tot}$ and $N_{cl}^{\rm tot}$ are the total number   
of local peaks in a SZE map and that of real clusters, respectively. 

\begin{figure}
\begin{center}\epsscale{1.0} 
\plotone{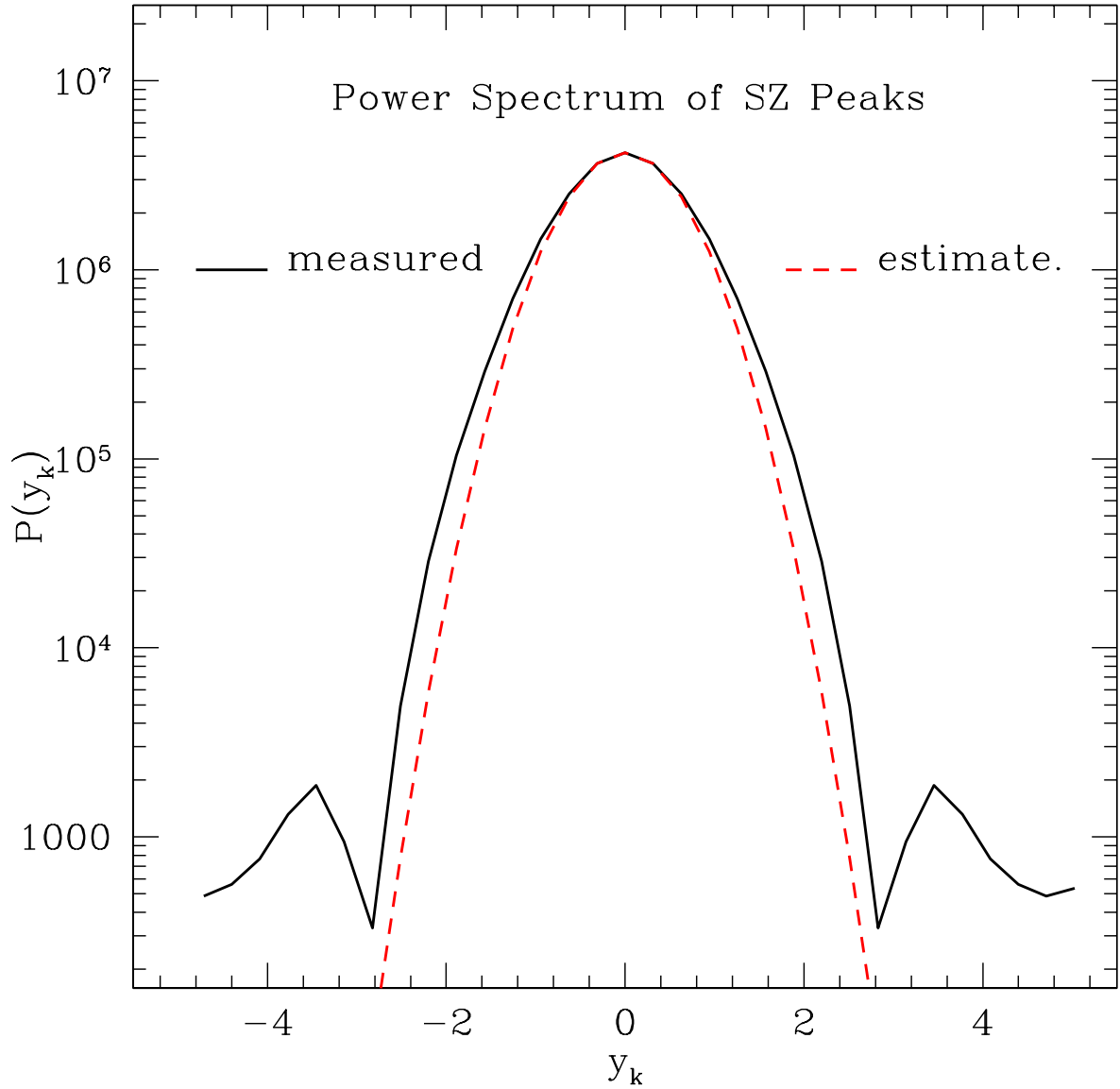}
\caption{The power spectrum of the rescaled $y$-peaks measured from one 
realization of the total SZE map. The solid line is the measured power 
spectrum, while the dashed line represents our estimate of the true power 
spectrum. 
\label{fig:power}}
\end{center}
\end{figure}

The stabilized deconvolution of $p(x)$ and $n^{(0)}_{cl}(y)$ can be 
performed in the Fourier space as 
\beq
\tilde{n}_{cl}(y_{k}) = 
\frac{W_{p}(y_{k})\tilde{n}^{(0)}_{cl}(y_{k})}{\tilde{p}(y_{k})},
\eeq
where $y_{k}$, $\tilde{n}_{cl}$, $\tilde{n}^{(0)}_{cl}$,
and $\tilde{p}$ are the Fourier counter parts of $y$,  
$n_{cl}$, $n^{(0)}_{cl}$, and $p$, respectively. Here, 
$W_{p}(y_{k})$ represents the Wiener optimal filter \citep{pre-etal92}, 
a key quantity to the deconvolution process. It is introduced 
to stabilize the process of the deconvolution on a discretized mesh. 
The functional form of $W_{p}(y_{k})$ is determined from the 
power spectrum of the $y$-peaks; the ratio of the true power spectrum 
($P_{-}$) to the measured one ($P_{+}$). The measured power spectrum
$P_{+}$ is usually a corrupted one due to the numerical noise effect. 
The true power spectrum $P_{-}$ is supposed to be well estimated by 
subtracting the obvious noise tail from the measured power spectrum 
$P_{+}$ (see Fig. \ref{fig:power}). Finally, the inverse Fourier
transform of $\tilde{n}_{cl}(y_{k})$ gives the desired estimate of the 
cluster number density $n_{cl}(y)$.  

The validity and stability of the above statistical method depends 
on the heights of $y$-peaks considering the definitions of $n^{(0)}_{cl}(y)$ 
and $p(x)$.  In the very low $y$ range where the noise peaks are dominant 
(say, $y < \sigma^{N}$, where $\sigma_{N}$ is the noise level), 
$n^{(0)}_{cl}(y)$ is close to zero, and hence the deconvolution likely fails. 
In other words, in the low $y$ range, most of the local peaks can
be safely regarded as just noise. Also, in the very high $y$ range (say, 
$y > 5\sigma_{N}$) where the signal peaks are dominant, we have 
$n^{(0)}_{cl}(y) \approx n_{sz}(y) \approx n_{cl}(y)$. 
In this range the noise contamination effect is almost completely
negligible so that the deconvolution of $p(x)$ is not meaningful any 
longer. Hence it is naturally expected that the above statistical 
method works best in the range of medium $y$ peaks 
(say, $\sigma_{N} < y \le 5\sigma_{N}$) where $n^{(0)}_{cl}(y)$ 
has non-zero values while containing a substantial 
noise-contamination that can be quantified as $p(x)$. 
We emphasize that it is this range of $\sigma_{N} < y \le 5\sigma_{N}$ 
where we hope to estimate the cluster number density more efficiently 
than the conventional method. 

\section{APPLICATION TO NUMERICAL SIMULATIONS}

\begin{figure}
\begin{center}
\epsscale{0.6} 
\plotone{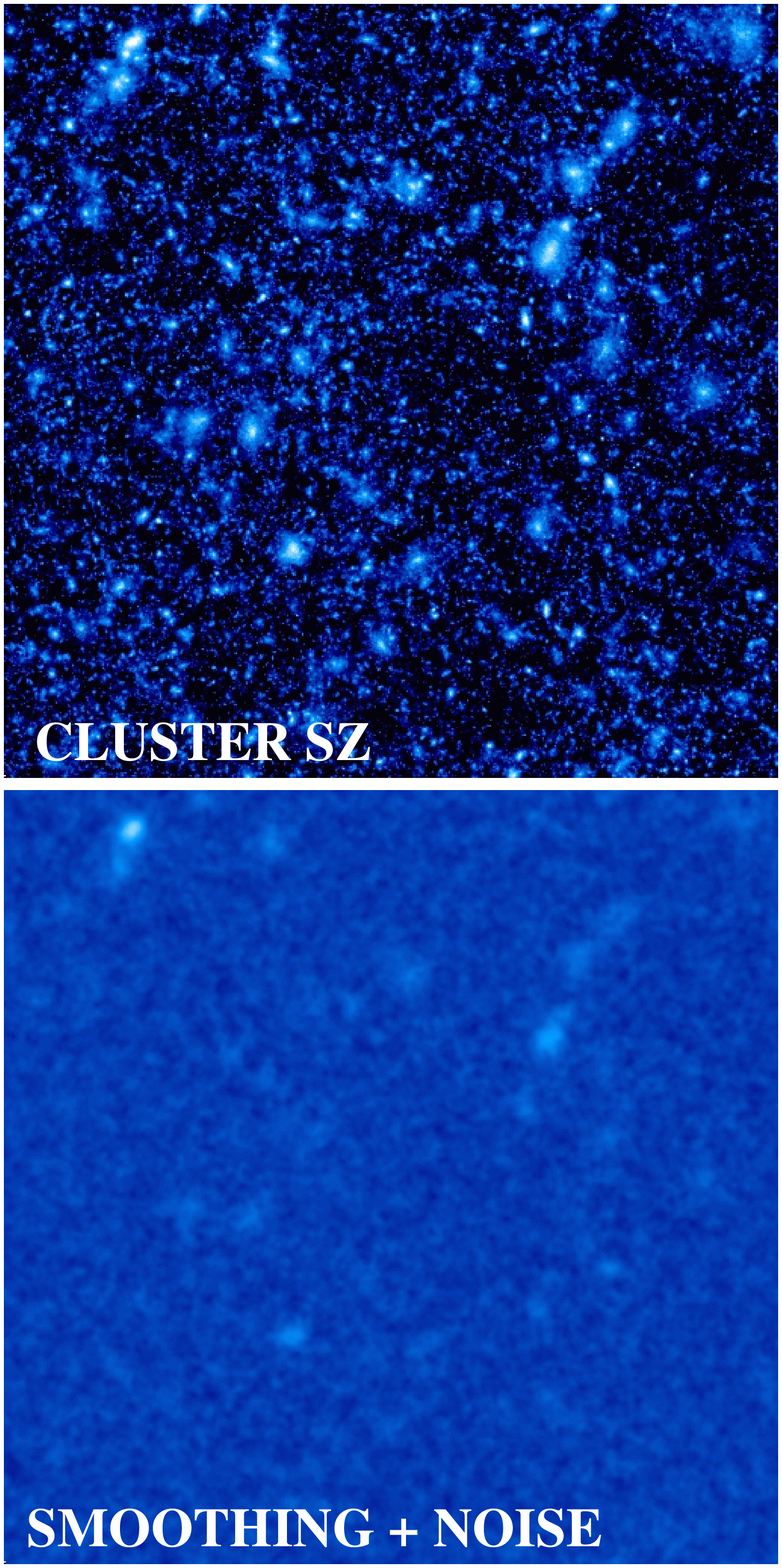}
\caption{Simulated thermal SZ effect maps. The maps shows 
the same field of view of 3.14 degree on a side. {\it Top}: a pure
thermal SZ effect; {\it Bottom}: a smoothed SZE map combined with 
a Gaussian noise field (see $\S 3$ in text).
\label{fig:omap}}
\end{center}
\end{figure}

We use the SZE cluster catalogues produced from a large cosmological
simulation carried out by the Virgo consortium
\citep{Jenkins01,yosh-etal01}. 
We make maps of $2048^{2}$ pixels in a field of view of 3.14 degree 
on a side. Details of the generation of the SZE cluster maps
are found in \citet{yosh02}.
For our purpose, we use a total of 8 realizations of the SZE cluster
maps,  each of which corresponds to a patch of the SZE sky map. 
We first smooth the SZE cluster maps by a Gaussian filter of 
scale radius $R_{g}$ in order to mimic the diffuse and extended clusters 
observed in real surveys. The scale radius $R_{g}$ corresponds to the
FWHM (full-width at half maximum) of the clusters which varies with 
different observations. Here we choose $R_{g} = 2$ arcminute, given that
 the natural beam is expected to have a FWHM of $2$ arcminute in AMiBA 
experiment. Next, we construct a Gaussian random field on the same 
$2048^{2}$ pixels using the Monte-Carlo method, assuming the flat-sky 
approximation. To simulate the instrumental noise of the real SZE 
observation, we adopt a white-noise power spectrum as the random field, 
and then convolve it with the noise-cleaning filter $W_{C}$ which is
given in Fourier $\bu$-space ($\bu$: the Fourier counterpart of $\bth$, 
$u \equiv \vert \bu \vert$) as \citep{pen-etal02,lee02}:
\begin{equation} 
W_{C}(u) = \frac{1}{u}\left[\exp\left(-\frac{u^{2}\theta^{2}_{A}}{2}\right) 
- \exp\left(-\frac{u^{2}\theta^{2}_{B}}{2}\right)\right], 
\end{equation}  
where $\theta_{B} = 6\theta_{A}$ with $\theta_{A} = 0.5$ arcminute 
suitable for AMiBA (Pen 2002, private communication). 
For each of the 8 realizations of SZE cluster maps, we construct an
independent noise field, and combine the cluster
sources with the noise field to construct a realistic, noise-added SZE map. 
In combining, we rescale each field by the y-dispersion of the noise, 
$\sigma_{y}^{N}$. Figure \ref{fig:omap} shows one realization of the sky 
maps of original SZE clusters (the top panel) and its smoothed version 
combined with the noise field (the bottom panel), respectively.  
From each realization of the total SZE map, we determine the number
density of the peaks $n_{sz}$ as a function of the rescaled peak
heights. Using the measured $n_{sz}$ along with equation 
(\ref{eqn:ng}), we compute the zero-th approximation to the cluster 
number density, $n^{(0)}_{cl}$, and measure the power spectrum of 
the peaks from the total SZE map in order to find a Wiener optimal 
filter $W_{p}$. Figure \ref{fig:power} shows an example of the peak 
power spectrum computed from one realization. 
The solid line corresponds to the measured power spectrum $P_{+}$ from 
the total SZE map realization, revealing the noise tail. 
The dashed line represents the estimate of the true power spectrum
$P_{-}$,  obtained by subtracting the noise tail from $P_{+}$. 
The Wiener optimal filter for the deconvolution is then approximated 
as $P_{-}/P_{+}$.  

\begin{figure}
\begin{center}\epsscale{1.0} 
\plotone{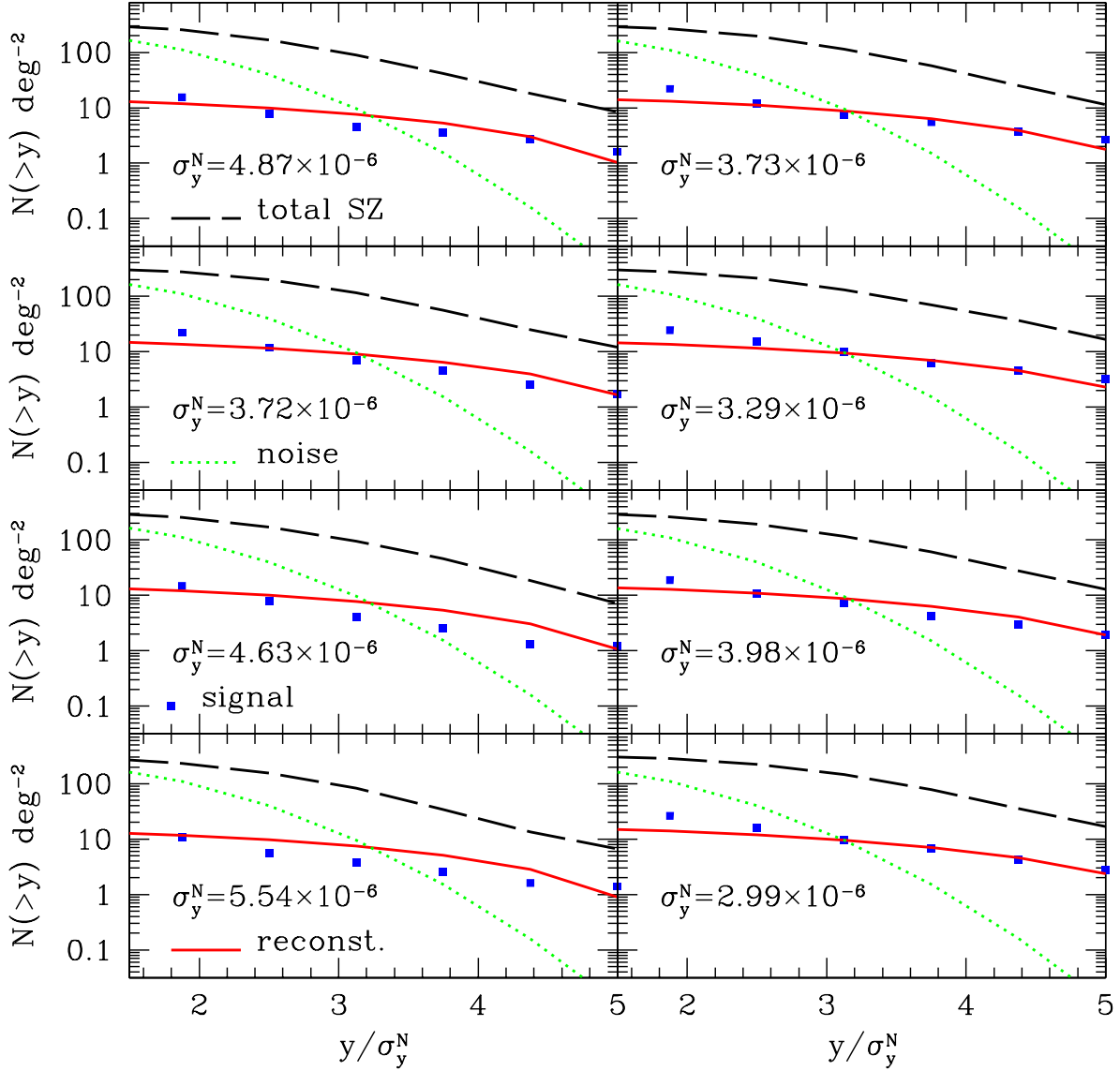}
\caption{The number counts of local peaks per degree square as a cumulative 
function of the rescaled Comptonization parameter in each realization.  
The long-dashed, the dotted, and the solid lines correspond 
to the number counts of the total peaks of the SZE maps, the noise,  
and the reconstructed signals, respectively, while the square dots 
correspond to the number counts of the true signals from the original 
SZE map from simulations before the combination with the noise field. 
\label{fig:recon}}
\end{center}
\end{figure}

\begin{figure}
\begin{center}\epsscale{1.0} 
\plotone{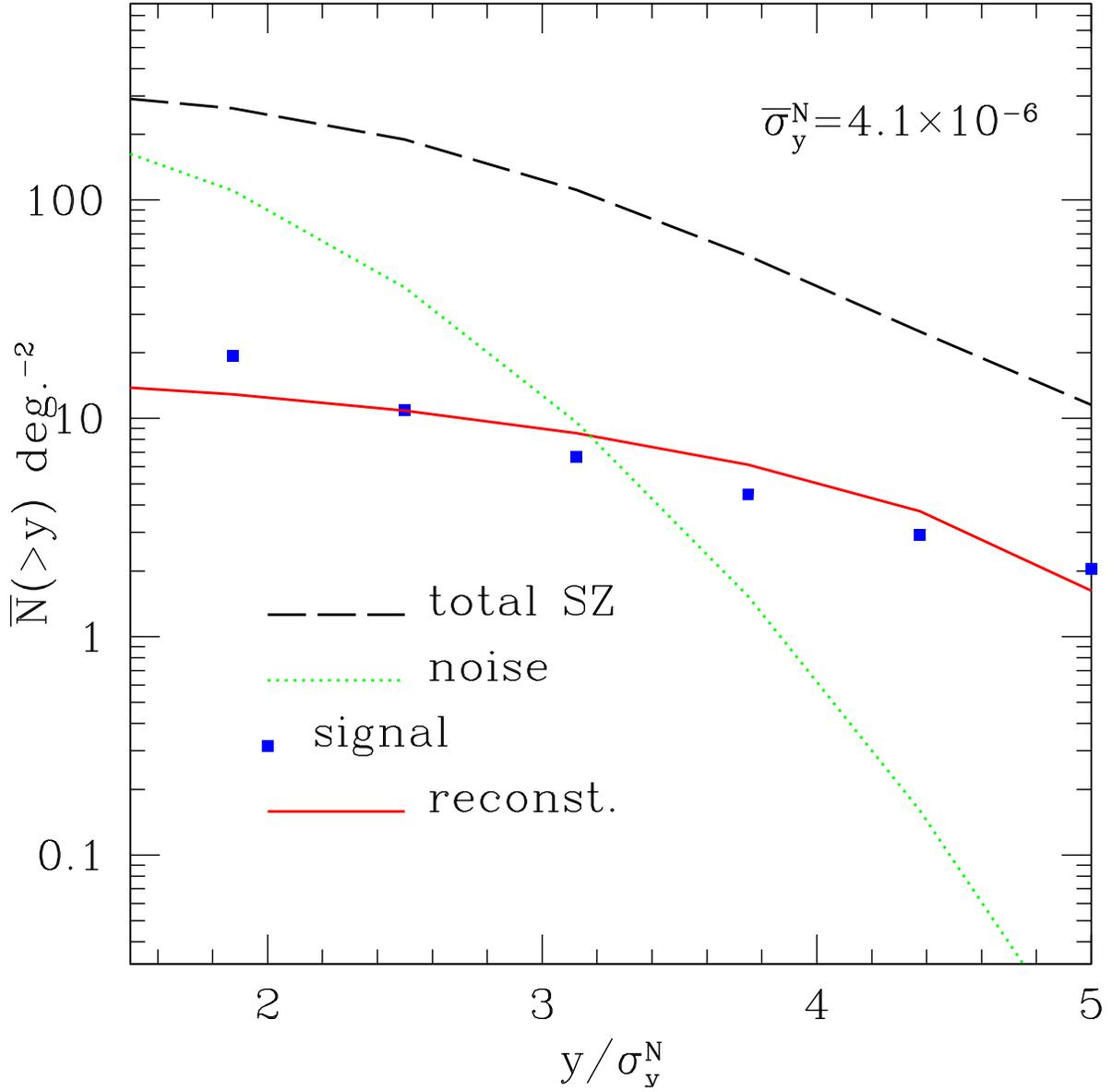}
\caption{Same as Figure \ref{fig:recon} but the averaged over the 
8 realizations.
\label{fig:trecon}}
\end{center}
\end{figure}

Finally, by using this Wiener optimal filter, we perform the 
stabilized deconvolution of the zero-th order signal distribution 
$n_{cl}^{(0)}$ and the one-dimensional Gaussian distribution $p(x)$ 
to estimate the true number density of the signal peaks. 
We found that for all realizations our method works quite well 
in the range of $\sigma_{y}^{N} < y \le 5\sigma_{y}^{N}$ 
within a $\sim 25 \%$ error. Figure \ref{fig:recon} plots the number 
counts of local peaks per degree square as a cumulative function of 
the rescaled Comptonization parameter 
$N_{cl}(\ge y) \equiv \int _{y}n_{cl}(y^{\prime}) dy^{\prime}$ for 
the case of each realization separately. 
The long-dashed, the dotted, and the solid lines represent 
the cumulative number densities of local peaks from the total SZE map, 
noise peaks from equation (\ref{eqn:ng}), and the signal peaks from 
the reconstruction, respectively, while the square dots represent the 
cumulative number densities of signals from the original SZE simulations 
before the noise contamination effect added.
Figure \ref{fig:trecon} plots the same as Figure \ref{fig:recon} but 
averaged over all 8 realizations. Our method reconstructs
also the {\it averaged} peak distribution fairly well.

\section{SUMMARY AND CONCLUSIONS}

We have tested the validity of the SZE statistics proposed by 
\citet{lee02} using numerical simulations. The statistical method is 
devised for the interferometric SZE observations in drift scanning mode, 
targeting at AMiBA as a model experiment, and applied to estimate 
the number density of cluster signals that occupy only 
a small fraction ($\sim 10 \%$) of the noise-contaminated 
SZE maps more efficiently than the conventional method based on 
the threshold cut-off. The method is developed to model the dominant 
noise contamination effect on the SZE peak distribution by a Gaussian 
scatter and to deconvolve the noise scatter and the SZE peak 
distribution. The stability of the deconvolution process has been 
achieved by convolution with a Wiener optimal filter estimated from the 
measurable power spectrum of the SZE peaks. 

We have applied the method to a large number of realistic SZE sky maps, 
and found that it indeed works for all realizations, giving a fairly 
good estimate of the cluster number density, including even those
clusters whose peak heights are compatible with noise-levels.   
In the conventional approach, those low-amplitude peaks are usually 
all disregarded as noise, so that a large fraction of low-amplitude cluster 
signals are missing. Our method allows us to count those missing low-amplitude
clusters as quickly and accurately as possible. Hence, we conclude that 
it will provide an efficient method to determine the number density
of galaxy clusters in the future interferometric SZE surveys. 

\acknowledgments

We thank K. Yoshikawa, U.L. Pen, and Y. Suto for helpful discussions 
and useful comments.  We both acknowledge gratefully the financial 
support of the JSPS (Japan Society of Promotion of Science)
fellowships.  This research was funded by the Grant-in-Aid for 
Scientific Research of JSPS (02674 and 12640231). 
The simulation data used in this paper are publicly available at 
www.mpa-garching.mpg.de/galform/virgo/vls

\clearpage

\clearpage

\clearpage

\clearpage

\clearpage

\end{document}